\documentclass[12pt]{article}
\usepackage[left=2cm, right=2cm, bottom=2cm ,top = 2cm, footskip=1cm]{geometry}
\usepackage[font=small,labelfont=bf]{caption}
\usepackage{xcolor}
\usepackage{setspace}
\usepackage{chngcntr}
\usepackage{array,multirow}
\usepackage{hyperref}
\usepackage{setspace}

\usepackage[authoryear,sort&compress]{natbib}
\setcitestyle{open={(},close={)}}
\usepackage{enumitem}
\usepackage{graphicx}
\usepackage{amsmath}
\usepackage{titlesec}
\titleformat{\paragraph}
{\normalfont\normalsize\bfseries}{\theparagraph}{1em}{}
\titlespacing*{\paragraph}
{0pt}{3.25ex plus 1ex minus .2ex}{1.5ex plus .2ex}

\usepackage[section]{placeins}

\title{Incomplete resection of the icEEG seizure onset zone is not associated with post-surgical outcomes}
\author{Sarah J. Gascoigne$^{1*}$, Nathan Evans$^{1}$, Gerard Hall$^1$, Csaba Kozma$^1$,\\  Mariella Panagiotopoulou$^1$, Gabrielle M. Schroeder$^1$, Callum Simpson$^1$, \\ 
Christopher Thornton$^1$, Frances Turner$^1$, Heather Woodhouse$^1$, \\ 
Jess Blickwedel$^1$, Fahmida Chowdhury$^3$, Beate Diehl$^3$, John S. Duncan$^3$,\\
Ryan Faulder$^1$, Rhys H. Thomas$^2$, Kevin Wilson$^4$,\\
Peter N. Taylor$^{1,2,3}$, Yujiang Wang$^{1,2,3**}$}

\begin{document}

\maketitle

\begin{enumerate}
\item{CNNP Lab (www.cnnp-lab.com), Interdisciplinary Computing and Complex BioSystems Group, School of Computing, Newcastle University, Newcastle upon Tyne, United Kingdom}
\item{Faculty of Medical Sciences, Newcastle University, United Kingdom}
\item{UCL Queen Square Institute of Neurology, Queen Square, London, United Kingdom}
\item{School of Mathematics, Statistics \& Physics, Newcastle University, United Kingdom}
\end{enumerate}

\begin{itemize}[label={}]
\item $^*$ Sarah Jane Gascoigne (ORCID ID: 0000-0003-1013-1875)\\
Urban Sciences Building, 
1 Science Square Newcastle upon Tyne, NE4 5TG\\
Tel: (+44) 191 208 4145 \hskip5em Email: S.Gascoigne@newcastle.ac.uk
\item $^{**}$ Yujiang Wang (ORCID ID: 0000-0002-4847-6273)\\
Urban Sciences Building, 
1 Science Square Newcastle upon Tyne, NE4 5TG\\
Tel: (+44) 191 208 4141  \hskip5em
Email: Yujiang.Wang@newcastle.ac.uk
\end{itemize}

\begin{center}
N.Evans5$^a$; Gerard.Hall$^a$; C.A.Kozma2$^a$; M.Panagiotopoulou2$^a$; Gabrielle.Schroeder$^a$; C.Simpson5$^a$; Chris.Thornton$^a$; Frances.Hutchings$^a$; H.Woodhouse$^a$; Jessica.Blickwedel$^a$;  Fahmidaamin.Chowdhury$^b$; B.Diehl$^c$; J.Duncan$^c$; Ryan.Faulder2$^b$;   Rhys.Thomas$^a$; Kevin.Wilson$^a$; Peter.Taylor$^a$

$^a$ @newcastle.ac.uk
$^b$ @nhs.net
$^c$ @ucl.ac.uk
\end{center}
\noindent{
We confirm that we have read the Journal’s position on issues involved in ethical publication and affirm that this report is consistent with those guidelines. \\
None of the authors has any conflict of interest to disclose.}
\newpage

\begin{doublespace}
\section*{Abstract}
\textbf{Objective:} Delineation of seizure onset regions from EEG is important for effective surgical workup. However, it is unknown if their complete resection is required for seizure freedom, or in other words, if post-surgical seizure recurrence is due to incomplete removal of the seizure onset regions.



\textbf{Methods:} Retrospective analysis of icEEG recordings from 63 subjects (735 seizures) identified seizure onset regions through visual inspection and algorithmic delineation. We analysed resection of onset regions and correlated this with post-surgical seizure control. 

\textbf{Results:} Most subjects had over half of onset regions resected (70.7\% and 60.5\% of subjects for visual and algorithmic methods, respectively). In investigating spatial extent of onset or resection, and presence of diffuse onsets, we found no substantial evidence of association with post-surgical seizure control (all $AUC<0.7$, $p>0.05$). 

\textbf{Significance:} Seizure onset regions tends to be at least partially resected, however a less complete resection is not associated with worse post-surgical outcome. We conclude that seizure recurrence after epilepsy surgery is not necessarily a result of failing to completely resect the seizure onset zone, as defined by icEEG. Other network mechanisms must be involved, which are not limited to seizure onset regions alone.

\subsection*{Key Points} 
\begin{itemize}
    \item Most regions involved in seizure onset tend to be resected in our adult icEEG cohort.
    \item A less complete resection is not associated with worse post-surgical outcomes.
    \item Onset and resection sizes are not significantly different across outcome groups.
\end{itemize}

\newpage
\section{Introduction} 
Where medications fail to control seizures in focal epilepsy, surgical resection is potentially curative \citep{cramer2021resective, lagarde2019repertoire, samuel2019seizure, west2019surgery}. Such surgeries aim to resect or disconnect epileptogenic tissue with the goal of post-surgical seizure freedom \citep{luders2001epilepsy}. Presurgical evaluation involves a battery of assessments across modalities which are used to localise tissue believed to be involved in epileptogenesis \citep{zijlmans2019changing}. More complex cases, for example those with no lesions visible on MRI, require intracranial electroencephalography (icEEG) recordings to localise epileptogenic tissue \citep{minotti2018indications}. The seizure onset zone (SOZ) is a proxy for the epileptogenic zone and is thus used to guide surgical resections \citep{rosenow2001presurgical} where appropriate.



With the concept of epilepsy as a network disorder, however, it is unclear if complete removal of seizure onset regions is necessary to attain seizure freedom post-surgically. Previous results have been contradictory. \cite{khan2022proportion} report that more complete resections of icEEG seizure onset channels is not associated with post-surgical seizure freedom in children, however numerous papers discuss `incomplete resection' as a mechanism of surgical failure \citep{vaugier2018role, englot2014factors}. 

To this end, we assessed the overlap between seizure onset regions and resected regions to determine if a larger proportion of onset regions were resected in patients with post-surgical seizure freedom or recurrence. Two methods were used to identify onset regions:  clinically labelled onset regions (CLOs), identified through visual inspection, and automatically labelled onset regions (ALOs), identified using a computational algorithm.
We also investigated if onset size and size of resection are associated with post-surgical seizure freedom.


\section{Methods} 

\subsection{Patient details}
In this retrospective study, 63 subjects were included with medically refractory focal epilepsy who underwent video-icEEG monitoring within the epilepsy monitoring units (EMU) at the National Hospital for Neurology and Neurosurgery (NHNN). 
All subjects had subsequent surgical resections, with seizure outcomes followed for at least one year. Surgical outcome groups did not differ in age, sex, or disease duration (See Suppl.~\ref{suppl:pat_meta}). 
Anonymised data were collected from NHNN; all analyses were completed following approval from the Newcastle University Ethics Committee (reference number: 28280/2022).




\subsection{EEG preprocessing} \label{preproc}
For each seizure recorded, we obtained icEEG recordings with 120 seconds of activity before onset and after offset. Prior to analyses, all data were resampled to 512~Hz.  An iterative noise detection algorithm was used to identify pre-ictal noise and validated by visual inspection. On a within-subject basis, channels identified as noisy were removed from all seizures (see Suppl. Methods~\ref{noisedetect}). Seizures with many noisy channels were removed to preserve the number of channels recorded per individual. When seizures occurred in close succession, we retained only the lead seizure. This resulted in 735 seizures being analysed.

Following the removal of noisy channels, recordings were re-referenced to a common average reference, notched filtered at 50Hz (and harmonics) with a 2Hz window to remove line noise, then band-pass filtered between 0.5 and 200Hz (fourth order, zero phase shift Butterworth filter).

\subsection{Clinically labelled onset (CLO)}
Seizure onset channels were labelled based on visual inspection of icEEG by an expert team at the NHNN. When multiple seizures occurred within the same subject, we obtained one CLO per subject as channels involved most frequently across clinically relevant focal seizure onsets, as indicated by the clinical team. 

\subsection{Automatically localised onset (ALO)}
\begin{figure}[h]
    \centering
    \includegraphics[scale=0.5]{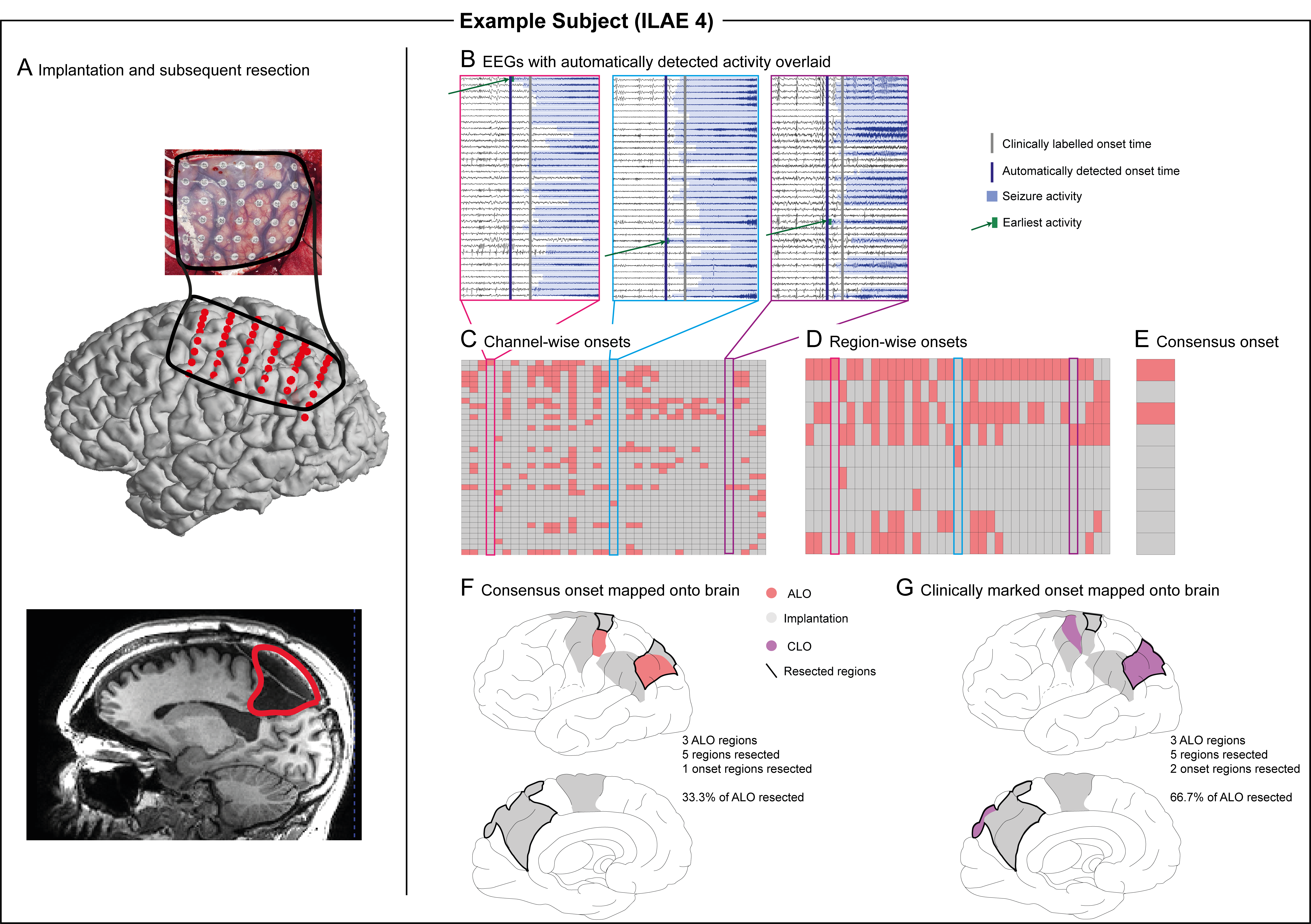}
    \caption{\textbf{Workflow from automatic seizure activity detection to onset regions for an example subject}. A: Electrode location and in-surgery image of implantation (top) and and post-surgical MRI with resected region outlined in red (bottom). B: icEEG time series for three focal seizures with detected activity highlighted in blue. First activity (i.e., onset) is highlighted in green and indicated with a green arrow. Clinically labelled and automatically detected onset time are labelled in  grey and blue, respectively. C: Channel-wise automatically detected seizure onsets. D: Region-wise automatically detected seizure onsets according to the Lausanne-scale 120 atlas. E: Region-wise consensus onset determined using regions involved in at least 50\% of onsets. F: Consensus onset (ALO) mapped onto cortex with onset highlighted in pink and recorded (non-onset) regions highlighted in grey. Additionally, the proportion of ALO resected is reported. G: Consensus onset mapped onto cortex with onset highlighted in purple and recorded (non-onset) regions highlighted in grey. Additionally, the proportion of CLO resected is reported. Note that resection here is labelled with respect to icEEG channel labels as registered to the Lausanne-scale 120 atlas. Gaps in resection reflect gaps in implantation rather than in the resection itself. }
    \label{fig:figure_1}
\end{figure}

Our seizure onset localisation extends the ``Imprint'' algorithm \citep{gascoigne2023library}. For each icEEG time series (See Fig. \ref{fig:figure_1}A for implantation and B for icEEG time-series) we computed eight markers of EEG activity level (line length, energy, and band powers in $\delta$, $\theta$, $\alpha$, $\beta$, low-$\gamma$, high-$\gamma$ bands) on a channel level for the preictal (120 seconds) and ictal periods using 1 second windows with 7/8 second overlap. 

Mahalanobis distance was employed to identify abnormal (pathological) activity across all eight markers, considering the covariance structure of markers. A baseline distribution of median absolute deviation (MAD) scores for Mahalanobis distances was established for each channel, with each window scored against all other time windows in the preictal segment. Abnormal MAD scores in the positive direction were considered, excluding points close to the median. Noise, such as interictal epileptiform discharges, was eliminated from the preictal segment based on abnormal MAD score. Subsequently, seizure activity in the ictal segment was identified by MAD scoring each window against the corresponding preictal distribution of Mahalanobis distances in the channel. Seizure activity was defined as MAD scores above 3 persisting for at least 80\% of a 9-second window. The first channel(s) with detected seizure activity, along with those within one second of onset, were labelled as seizure onset. Channel-wise onsets (Fig. \ref{fig:figure_1}C) were then localised to regions of interest (referred to as regions in this work) according to the Lausanne-scale 120 atlas \citep{hagmann2008mapping} (Fig. \ref{fig:figure_1}D). To summarise onset channels across seizures in a given patient, we retained any channels that were involved in at least 50\% of all recorded focal seizures (Fig. \ref{fig:figure_1}E \& F). Similarly, each subject's clinically labelled onset channels were localised to Lausanne-120 regions (Fig. \ref{fig:figure_1}G).

\subsection{MRI processing for identifying regions and resected tissue}
To map electrode coordinates to brain regions we used the same methods as described previously \citep{wang2023temporal}. In brief, we assigned electrodes to regions from the Lausanne-scale 120 atlas. We used FreeSurfer to generate volumetric parcellations of each patient's pre-operative MRI \citep{hagmann2008mapping, fischl2012freesurfer}. Each electrode contact was assigned to the closest grey matter volumetric region within 5~mm. If the closest grey matter region was $>$5mm away then the contact was excluded from further analysis. 

All onset channels (CLO and ALO) were subsequently mapped to these atlas regions to allow comparison between subjects in terms of number/percentage of regions with onset or resected. A region was considered to be part of onset, if at least one channel in the region was in onset. 

To identify which regions were later resected, we used previously described methods \citep{taylor2018impact, taylor2022normative}. We registered post-operative MRI to the pre-operative MRI and manually delineated the resection cavity. This manual delineation accounted for post-operative brain shift and sagging into the resection cavity. Electrode contacts within 5~mm of the resection were assigned as resected. Regions with $>$25\% of their electrode contacts removed were considered as resected for downstream analysis (see \cite{taylor2022normative}). 

\subsection{Comparing the proportion of onset resected, onset size, and resection size across outcome groups}
We labelled ILAE 1-2 as having a `favourable' outcome as they did not experience post-surgical seizures, ILAE 3+ were labelled as having `unfavourable' outcomes.
We computed the proportion of the onset that was subsequently resected. Regions in the ALO and CLO were directly compared against resection (Fig. \ref{fig:figure_1} F \& G). Proportion of onset resected, number of regions in onset, and number of regions in resection were examined across all subjects and were compared across outcome groups using logistic regression with the following formula:
$$ln(\frac{P}{1-P})\sim variable$$
where $ln$ is the natural log, $P$ is the probability of a subject being in the unfavourable outcomes category (ILAE 3+), and $variable$ is the subject-level value computed for each comparison (e.g., proportion of onset resected). We computed the area under the receiver operating characteristic curve (AUC) for the model and the associated p-value using a permutation test with 1000 permutations. AUC thresholds reported follow previous conventions \citep{mandrekar2010receiver}: AUC$\geq 0.7$ is acceptable and AUC$\geq 0.8$ is excellent. Any AUC$<0.7$ was considered unacceptable (i.e. the model is not able to distinguish outcome groups). 



\subsection{Data and code availability}
The analysis code and data are available on Zenodo.org upon acceptance. 
 
\section{Results} 



\subsection{icEEG Seizure onset regions tend to be resected, but more complete resections are not associated with more favourable surgical outcomes}
First, we investigated if seizure onset regions tend to be resected, and found this to be the case across all patients. The median proportion of clinically labelled onset (CLO) regions resected and automatically labelled onset (ALO) regions resected was 80\% (100\% ILAE 1-2, 77.75\% ILAE 3+) and 75\% (69.05\% ILAE 1-2, 85.71\% ILAE 3+), respectively. Most CLO and ALO regions were resected in 70.73\% (29/41) and 60.47\% (26/43) of subjects, respectively. 

\begin{figure}
     \centering
     \includegraphics[scale=0.75]{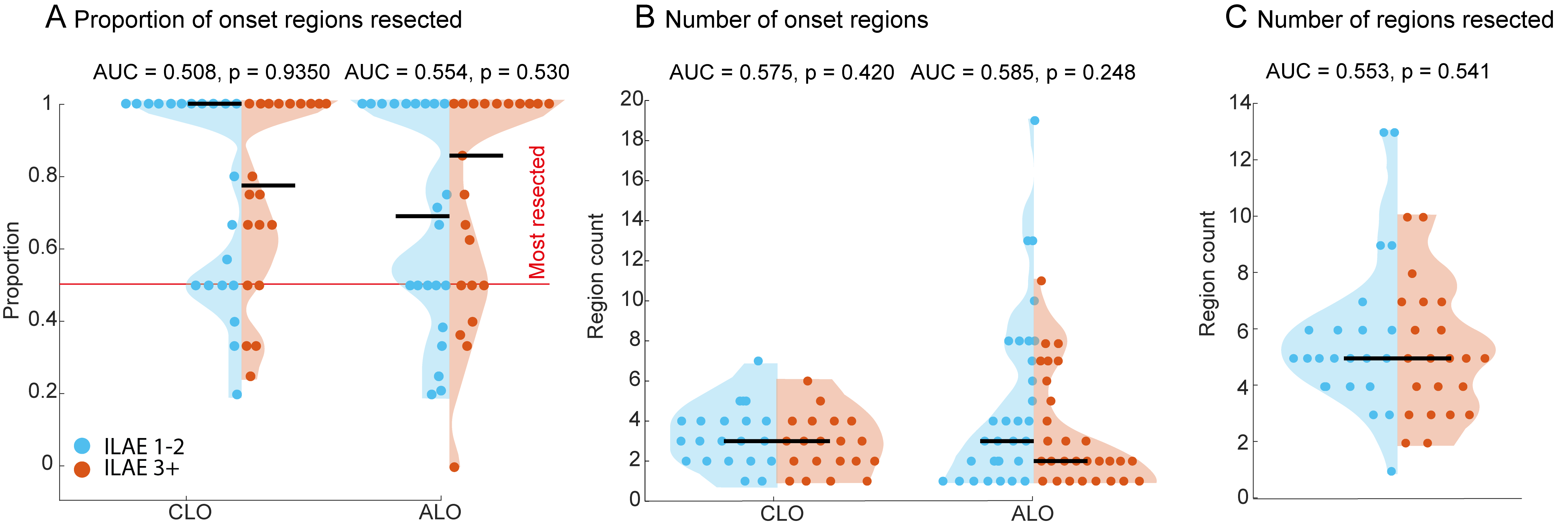}
     \caption{\textbf{Comparing proportion of onset resected and counts of onset and resected regions across ILAE outcome groups}. `Favourable' (ILAE 1-2) and `unfavourable' (ILAE 3+) surgical outcomes are displayed in blue and orange, respectively. A: Proportion of clinically labelled (CLO) and automatically labelled (ALO) onset regions resected across outcome groups are displayed on the left and right, respectively. B: Count of Lausanne-120 regions in clinically labelled (CLO) and automatically labelled (ALO) onsets across outcome groups are displayed on the left and right, respectively. C: Count of Lausanne-120 regions resected across outcome groups.}
     \label{fig:figure_3}
 \end{figure}

There was no difference in the distributions of the proportion of onset regions resected between subjects with favorable (ILAE 1-2) and unfavorable outcomes (ILAE 3+) for both CLO and ALO ($AUC<0.7$, see Figure \ref{fig:figure_3}A).  

\subsection{Larger onsets and resections are not associated with surgical outcomes}
We used the number of onset regions in the Lausanne-120 atlas as a proxy for onset size. For CLO and ALO, the median onset size was 3 regions (3 ILAE 1-2, 3 ILAE 3+) and 3 regions (3 ILAE 1-2, 2 ILAE 3+), respectively. Figure \ref{fig:figure_3}B displays onset sizes across outcome groups and onset detection methods with corresponding AUCs and p-values. 

Similarly, the median number of regions resected with icEEG sampling was 5 regions (5 ILAE 1-2, 5 ILAE 3+).  Figure \ref{fig:figure_3}C displays number of regions across outcome groups with corresponding AUCs and p-values. In all comparisons the AUC was below 0.7 with corresponding p-values greater than 0.05.

 
\section{Discussion} 
We observed that a more complete resection of the observed icEEG contacts associated with seizure onset and early spread, both clinically labelled and automatically detected, was not associated with surgical outcomes. Interestingly, the median proportion of onset regions resected was greater in those with unfavourable outcomes compared to favourable outcomes. Potential confounding variables of onset and resection sizes were not found to differ across surgical outcome groups. The results of this work support the conclusion that surgical failure is not necessarily a result of an incomplete resection of the seizure onset zone as defined by icEEG. This is in line with the idea that post-surgical seizure freedom/relapse may be driven by network mechanisms beyond the seizure onset regions \citep{carboni2019network, corona2023non}. Alternatively, our findings may reflect that the `true' SOZ is not covered by icEEG for all subjects. 

This work has a large sample size given the type of study. However, the sampled subjects are highly heterogeneous and do not represent a general epilepsy population. Our results are perhaps more specific to a difficult-to-treat cohort, given that these patients required intracranial monitoring. Future work could incorporate data from multiple sites, including both adult and paediatric cases. Potential mediating factors, for example age and surgery/recording sites, should be investigated alongside icEEG-based seizure onset markers. We are, however, encouraged by findings similar to ours in a pediatric cohort \citep{khan2022proportion}.

The ALO algorithm uses a preictal baseline from which ictal activity is detected. This offers a limited view on potential baseline activity and assumes that there are no key preictal changes in activity. However, it is important to note that icEEG activity fluctuates in time \citep{panagiotopoulou2022fluctuations,wang2023temporal}, this means that a baseline from immediately preceding activity is able to account for such fluctuations. Therefore, results are less likely to be skewed by such effects, and this approach is also most similar to the visual inspection approach taken by the clinical team. 


Intracranial recordings have limited spatial coverage, therefore it is possible that all seizure onset/epileptogenic regions were not captured in all subjects. Future work could incorporate non-invasive techniques with whole-cortex coverage to capture potential abnormalities outside the recorded regions \citep{horsley2023complementary, owen2023meg}. We conclude that a multi-modal approach may offer more information pertaining to mechanisms behind post-surgical seizure recurrence in epilepsy surgery. In this work, we hypothesised that resecting a greater proportion of the seizure onset zone would result in more favourable outcomes, however our results do not provide evidence in support of this hypothesis.

In conclusion, the results of this work demonstrate that seizure onset, both clinically labelled and automatically detected, tends to be resected. However, contrary to expectations, resecting a larger proportion of the onset is not associated with surgical outcomes. This means that, in validating markers of epileptogenic brain regions, resection of the onset alone may not be a suitable ground truth.

\end{doublespace}
\newpage

\bibliography{refs}

\newpage


\renewcommand{\thefigure}{S\arabic{figure}}
\renewcommand{\thetable}{S\arabic{table}} 
\counterwithin{figure}{subsection}
\counterwithin{table}{subsection}
\renewcommand\thesubsection{S\arabic{subsection}}
\setcounter{subsection}{0}

\section*{Supplementary}
\subsection{Glossary of acronyms}
\textbf{Clinical terms}
\begin{itemize}[label={}]
    \item \textbf{ASMs:} Anti-seizure medications
    \item \textbf{EEG:} Electroencephalography
    \item \textbf{EMU:} Epilepsy Monitoring Unit
    \item \textbf{FLE:} Frontal lobe epilepsy
    \item \textbf{icEEG:} Intracranial EEG
    \item \textbf{ILAE:} International League Against Epilepsy
    \item \textbf{ROI:} Region of interest
    \item \textbf{TLE:} Temporal lobe epilepsy
\end{itemize}

\textbf{Statistical terms}
\begin{itemize}[label={}]
    \item \textbf{CAR:} Common average reference
    \item \textbf{MAD:} Median absolute deviation
\end{itemize}

\subsection{Subject Metadata\label{suppl:pat_meta}} 
We investigated underlying associations between metadata and surgical outcome to ensure that our results were robust. Here we report surgical outcomes (Table \ref{supl_tab:ilae_out}), age, epilepsy duration (Table \ref{supl_tab:cont_meta}), sex, and epilepsy surgery type (Table \ref{supl_tab:cat_meta}). For each variable, we tested if there was a significant effect on surgical outcome using independent samples t-tests for continuous data and $\chi^2$ tests for categorical data. Any significant effects ($p<0.05$) would be regressed out of our data, however no such effects existed therefore we were confident to continue with our analyses as planned. 

\begin{table}[]
    \centering
    \begin{tabular}{lccc|c}
        \hline
        &All&ILAE 1 -2& ILAE 3+&$|t|(df)$\\
        \hline
         Age&33.89&32.94&34.93&$|t|(61) = -0.780$  \\
         &(10.10)&(10.47)&(9.74)&$p=0.438$\\
         \hline
         Epilepsy duration&20.75&22.03&19.33&$|t|(61) = 1.04 $ \\ 
         &(10.24)&(10.04)&(10.44)&$p=0.300$\\
         \hline
    \end{tabular}
    \caption{Table reporting continuous metadata across subjects. Data are represented as mean(standard deviation). T test-statistics and associated p-values are reported to capture if there is a significant association between metadata categories and surgical outcomes.}
    \label{supl_tab:cont_meta}
\end{table}

\begin{table}[]
    \centering
     \begin{tabular}{lrccc|c}
        \hline
        &&All&ILAE 1 -2& ILAE 3+&$\chi^2(df,N)$\\
        \hline
        Sex&M&30&14&16&$\chi^2(1,63)=0.7498$\\
        &   F&&33&19&14$p=0.387$\\
        \hline
        Surgery type&T Lx&28&13&15&$\chi^2(2,63)=0.928$ \\
        &            F Lx&27&16&11& $p=0.629$\\
        &            Other&8&4&4&\\
        \hline
    \end{tabular}
    \caption{Table reporting categorical metadata across subjects. Counts are reported for each group. $\chi^2$ test-statistics and associated p-values are reported to capture if there is a significant association between metadata categories and surgical outcomes.}
    \label{supl_tab:cat_meta}
\end{table}

 We labelled ILAE 1-2 as having a`favourable' outcome as they did not experience post-surgical seizures, ILAE 3+ were labelled as having `unfavourable' outcomes. Table \ref{supl_tab:ilae_out} lists all ILAE outcomes, the count of subjects in each group, and a definition of the clinical observations required for a subject to be included in each group. 

\begin{table}[]
    \centering
    \begin{tabular}{|l c c  p{8cm}|}
        \hline
         Category&ILAE &Count & Definition  \\
         &outcome&(n=63)&\\
         \hline
        Favorable & 1 &25&Completely seizure free; no auras \\
        &2&8&Only auras; no other seizures\\
        \hline
        Unfavorable &3& 10&1 to 3 seizure days per year; $\pm$ auras \\
        &4&14&4 seizure days per year to 50\% reduction of baseline seizure days; $\pm$ auras\\
        &5&6& Less than 50\% reduction of baseline seizure days; $\pm$ auras\\
        \hline
    \end{tabular}
    \caption{Table of the counts of ILAE outcomes across subjects. We labelled ILAE 1-2 as having a`favourable' outcome as they did not experience post-surgical seizures, ILAE 3+ were labelled as having `unfavourable' outcomes. Here we have included the definitions for each ILAE surgical outcome group based on \cite{commission2001proposal} as reference, for clarity.}
    \label{supl_tab:ilae_out}
\end{table}

\subsection{Supplementary Methods}
\subsubsection{Noise Detection} \label{noisedetect}
Prior to onset detection, the icEEG time series of each seizure segment was assessed for noise.  Line noise was removed using a notch filter at 50Hz and 100Hz (with 2Hz windows). As the preictal segment was used in the computation of the imprint, an iterative noise detection algorithm was applied to identify noise in the preictal segment as follows: 
\begin{enumerate}
    \item Raw icEEG time series MAD scored based on variance and min-max range for each channel independently 
	\item MAD$>$16 labelled as ‘outlier’ - channel is noisy 
	\item ‘Noisy’ channels removed 
	\item icEEG time series common average referenced (CAR)  
	\item MAD$>$16 labelled as ‘outlier’ - channel is noisy 
	\item ‘Noisy’ channels removed 
	\item 1Hz high-pass Butterworth $4^{th}$ order filter used to remove any slow trends 
	\item Repeat the process with a less lenient threshold of MAD$>$12.  
	\item Visual check  
\end{enumerate}
Visual checks of time series and power spectral densities (PSD) plots were performed to identify any noise in the preictal segment not detected in our algorithm. Further, noise in the ictal segment was visually assessed using icEEG traces and PSD plots - noisy channels were removed from all recordings. We did not seek or remove noise impacting only the postictal segment. Following this, the data was processed, as described in Section \ref{preproc}.

\subsubsection{Preictal artifact removal} \label{suppl:preict_artifact}
To ensure that our baseline activity was not skewed by artifacts (e.g., noise or interictal spikes), we used an extension of the imprint algorithm to identify and remove outliers from the baseline distribution. For each channel, we computed the Mahalanobis distance for each time point compared to all time windows in the preictal period. We then used a MAD scores with a threshold of 3 to highlight outliers. We removed time windows highlighted, replacing all features with NaN in the feature matrix, as abnormal then recomputed Mahalanobis distances and MAD scores for the preictal baseline from which the ictal period could be scored. 

\begin{figure}
    \centering
    \includegraphics{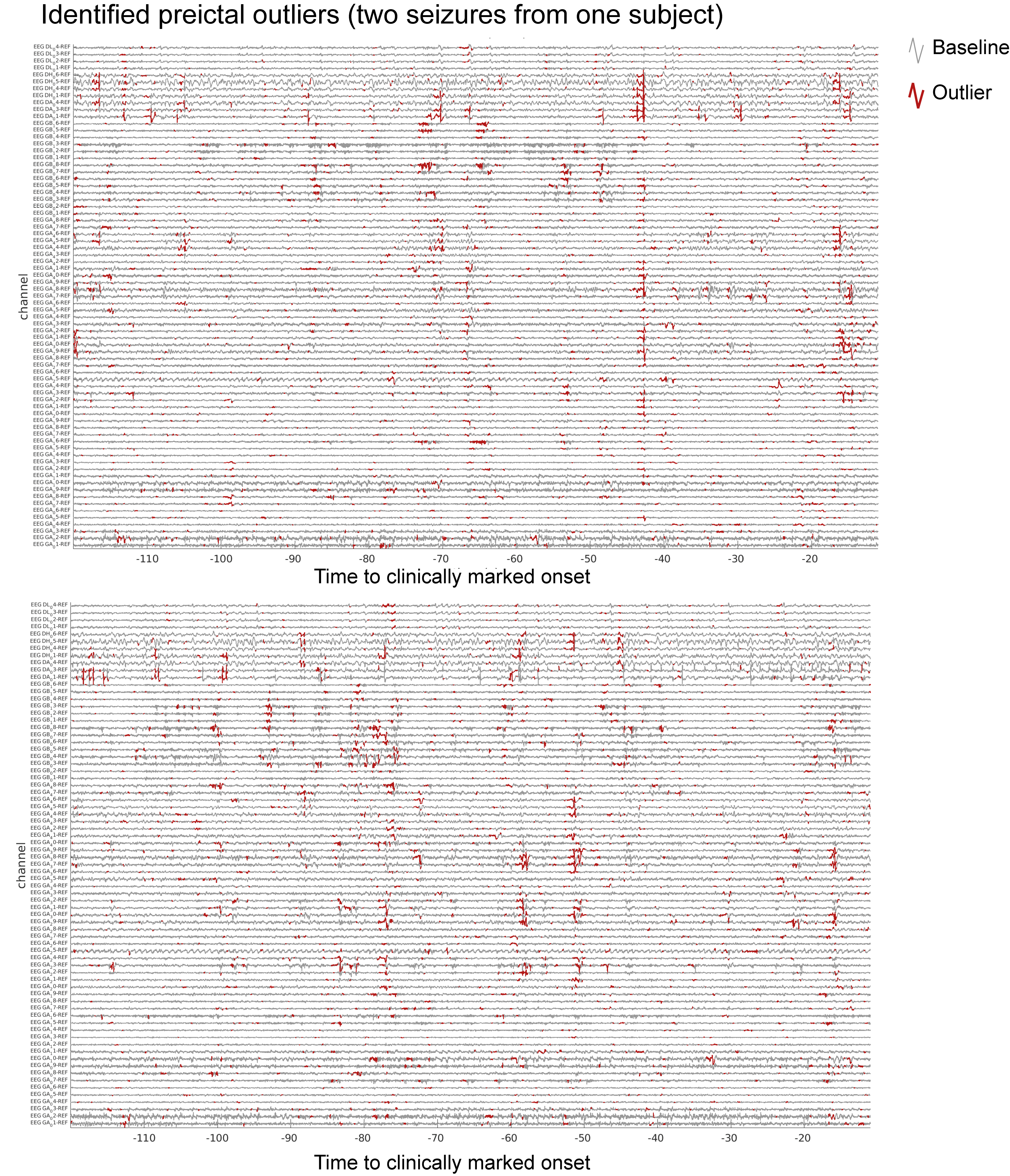}
    \caption{Two preictal periods from the same subject with baseline activity shown in grey and artifacts (MAD of Mahalanobis distances $>3$) shown in red. Time windows shown in red were removed from 3-dimensional marker matrices then Mahalanobis distances and MAD scores for the preictal baseline were recomputed. }
    \label{fig:enter-label}
\end{figure}

\subsection{Supplementary Results}
Here we have investigated associations between surgery type and concordance between CLO and ALO for subjects with temporal or frontal lobe resections. In this way, we can rule out potential skewing of results due to epilepsy type.  
\subsubsection{Surgical outcome is not associated with surgery type}
Across the 55 subjects with temporal or frontal lobe resections (see Table \ref{supl_tab:cat_meta}), we found no significant association between surgery type and outcome group ($\chi^2(1,55) = 0.9078, p=0.341$). There were more `favourable' outcomes than `unfavourable' outcomes in frontal lobe resections and vice versa for temporal lobe resections; however, the differences in group size do not reach significance. This finding somewhat contradicts the findings in \cite{meng2023multivariate} which suggested that a temporal lobe resection is a positive predictor for post-surgical seizure freedom. However, the lack of significance and small sample size in this study mean that this is likely a result of chance and not a cause for concern.

\subsubsection{ALO is not highly concordant with CLO}
We computed Cohen's $\kappa$ between the clinically labelled onset and the consensus onset. We further compared each seizure onset with the CLO and captured the maximum concordance across seizures. 
We additionally compared outcome groups and found that a logistic regression model was not able to adequately distinguish between the two groups (see Suppl. Fig. \ref{suppl_fig:comp_parcs}). Therefore, we have no evidence that differences between clinically and automatically labelled onsets are suggestive of worse surgical outcomes. 

\begin{figure}
    \centering
    \includegraphics[width=0.5\textwidth]{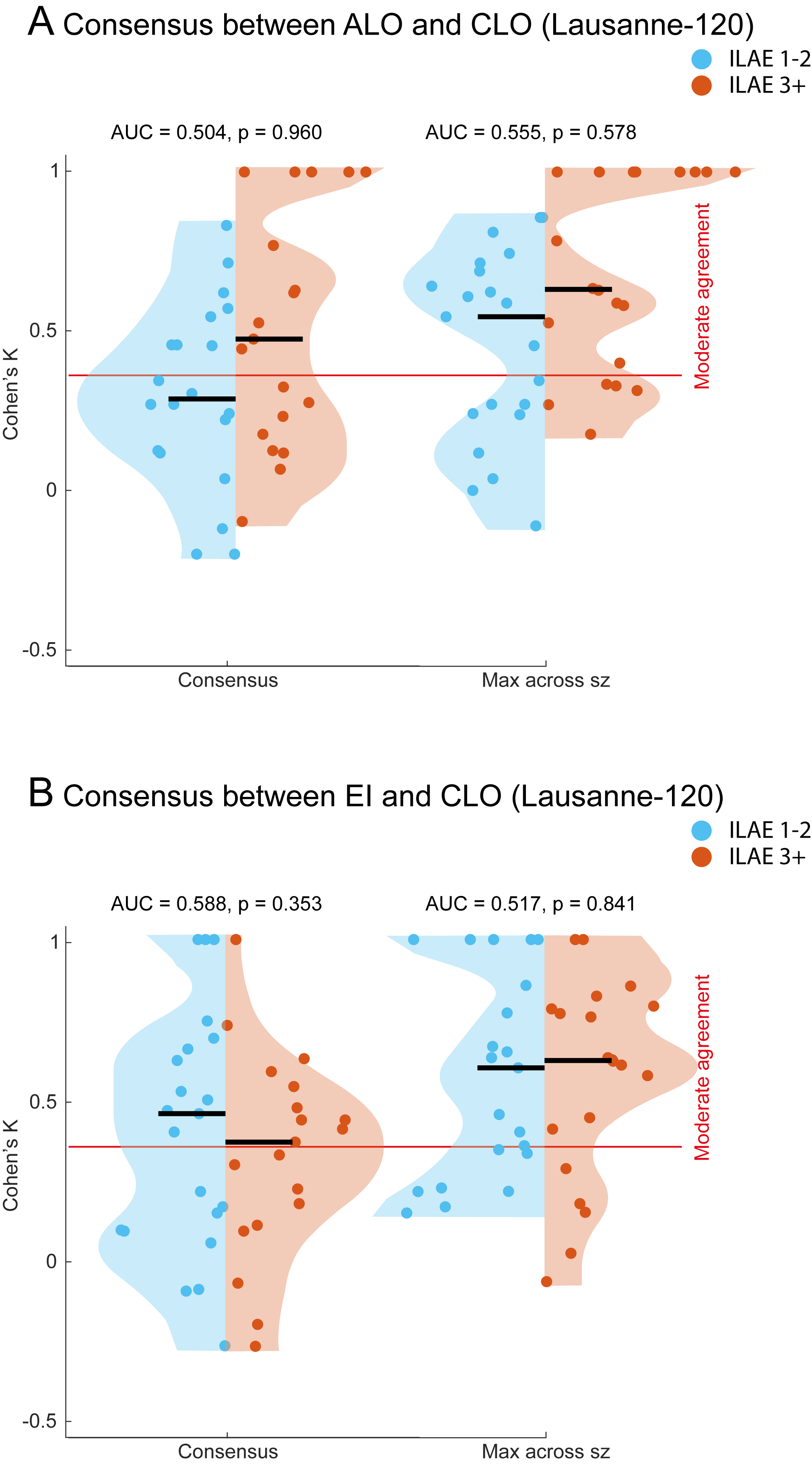}
    \caption{\textbf{Violin plots showing concordance across outcome groups (ILAE 1-2 in blue and ILAE 3+ is orange). `Moderate concordance (i.e. Cohen's $\kappa > 0.4$) lies above the red line.} A: Concordance between CLO and consensus onset (left) and maximum concordance between CLO and any one seizure (right) across outcome groups. B: Onsets based on EI rather than imprint algorithm. Concordance between CLO and consensus onset (left) and maximum concordance between CLO and any one seizure (right) across outcome groups.}
    \label{suppl_fig:clo_concord}
\end{figure}

\subsubsection{Results are consistent across parcellations} \label{suppl:across_parc_schemes}
\begin{figure}
    \centering
    \includegraphics[width=0.7\textwidth]{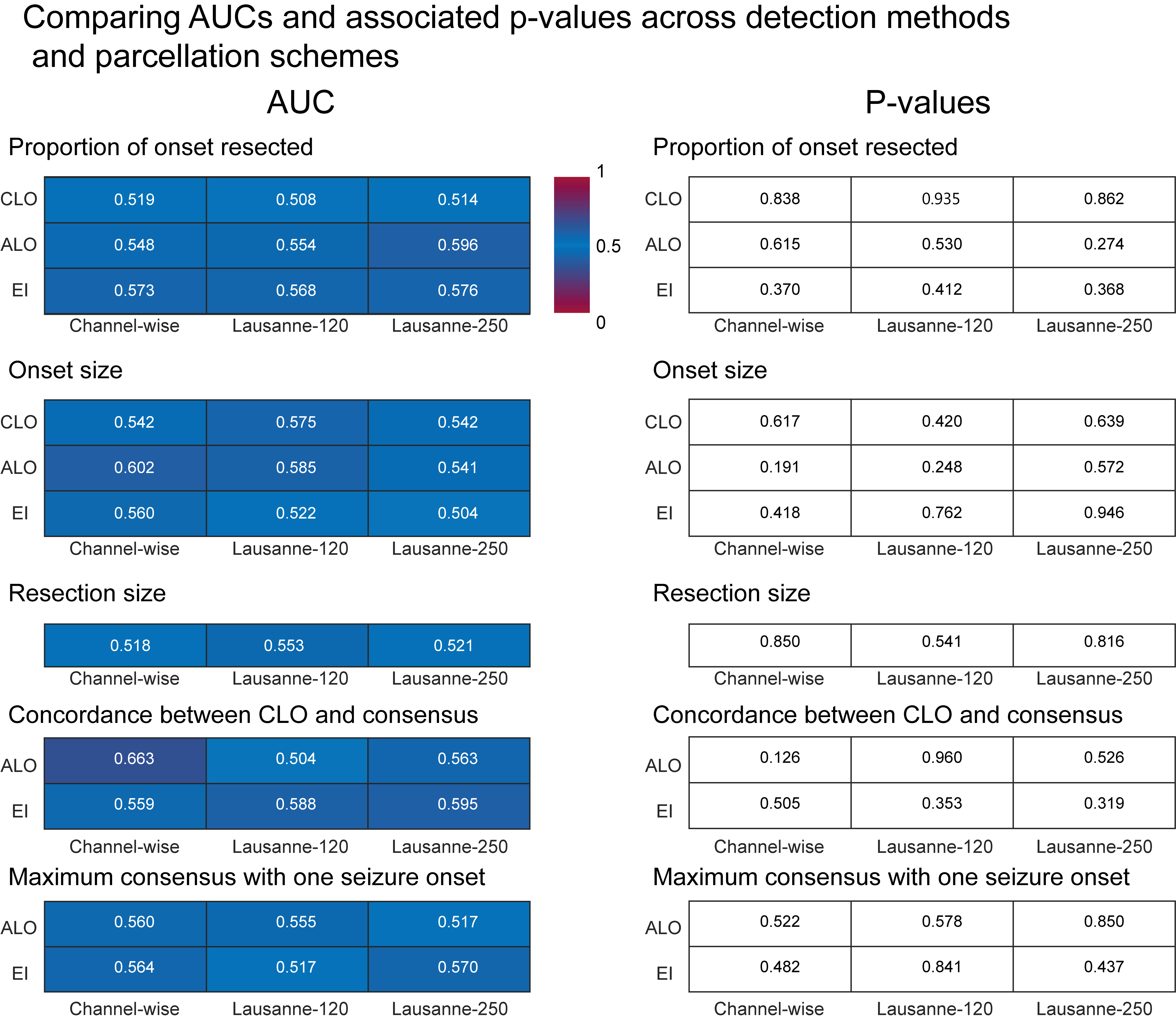}
    \caption{Comparing AUC and associated p-values across onset detection methods and parcellation schemes. P-values below 0.05 (of which there are none) are highlighted in red.}
    \label{suppl_fig:comp_parcs}
\end{figure}
All analyses performed within this work were repeated using channel-wise onsets and ROI-wise onsets using the Lausanne-250 atlas. To ensure that results were not simply a result of the selected parcellation scheme, we compared AUC values and associated p-values across parcellations. We additionally included the Epileptogenicity Index (EI) \citep{bartolomei2008epileptogenicity} to confirm that our ALO results were not exclusive to the algorithm used in this work. Supplementary figure \ref{suppl_fig:comp_parcs} displays heat-maps of AUC values across parcellation schemes and onset detection methods (left) and associated p-values (right). Clearly, there are no notable shifts in how well outcome groups can be distinguished, irrespective of the comparison. All AUCs were consistently below 0.7, with only slight variations across parcellation schemes (Suppl. Fig. \ref{suppl_fig:comp_parcs}A).
All p-values were greater than 0.05 (Suppl. Fig. \ref{suppl_fig:comp_parcs}B), further supporting that outcome groups do not significantly differ in proportion of onset resected, onset size, resection size, and even concordance between automatically detected and clinically labelled onsets. 

\begin{figure}
    \centering
    \includegraphics[width=0.5\textwidth]{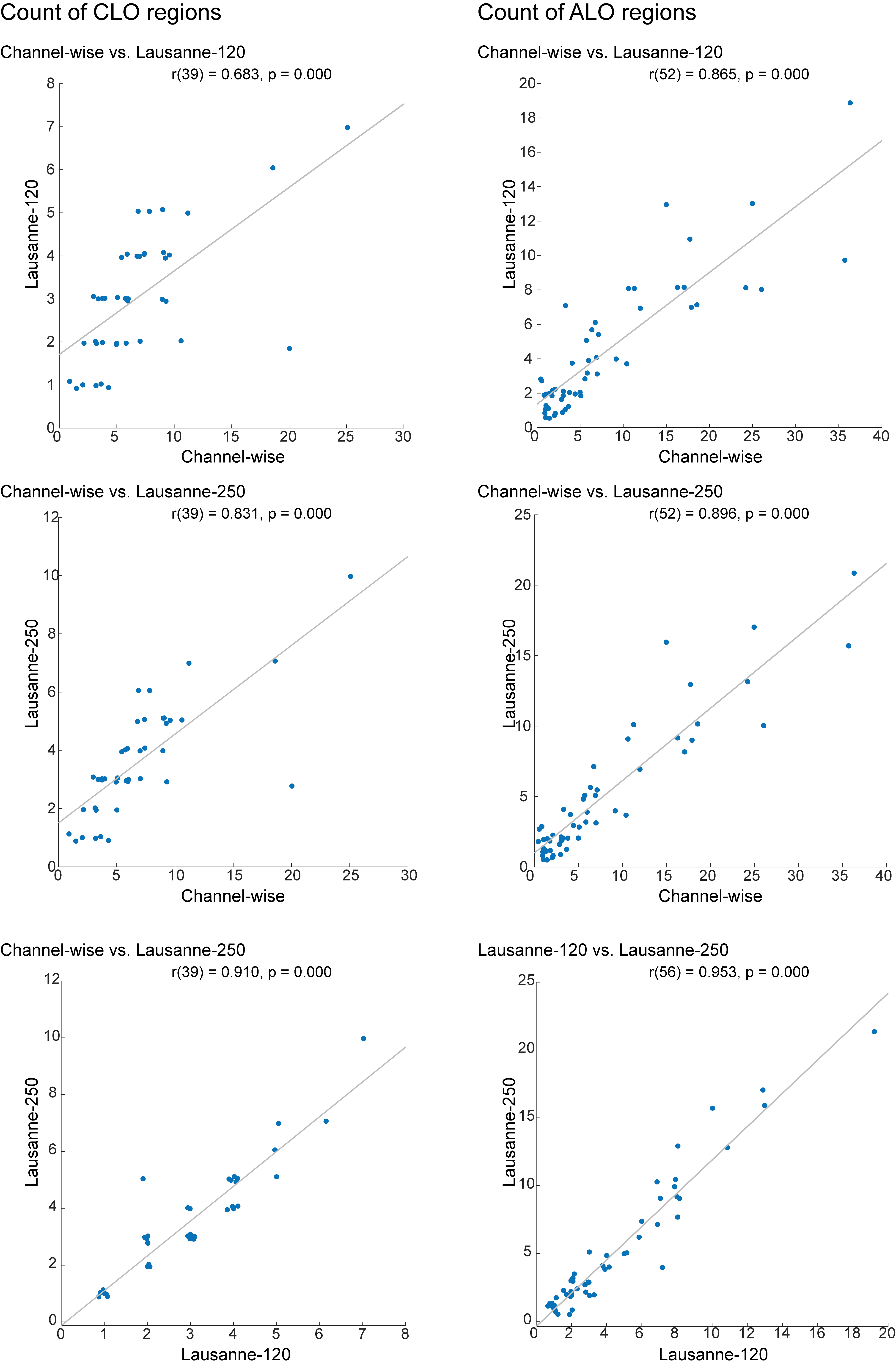}
    \caption{\textbf{Scatterplots of subject-wise counts of onset regions across channel-wise, Lausanne-120, and Lausanne-250 parcellations.} Left: count of clinically labelled onset (CLO) regions. Right: count of automatically labelled onset (ALO) regions. Top: comparison of channel-wise and Lausanne-120 onsets. Middle: comparison of channel-wise and Lausanne-250 onsets. Bottom: comparison of Lausanne-120 and Lausanne-250 onsets.}
    \label{suppl_fig:comp_parc_ons_size}
\end{figure}

We also checked the subject-level changes across parcellation schemes and onset detection methods using correlation analyses (see Suppl. Fig. \ref{suppl_fig:comp_parc_ons_size}). Here we plotted the count of CLO and ALO regions using channels, Lausanne-120 regions, or Lausanne-250 regions. The range of values differed across parcellations, as expected. Most Spearman's rank correlation coefficients ($\rho$) were greater than 0.8, $\rho$ = 0.68 when comparing the count of CLO regions in channels and Lausanne regions. In all cases $p<0.001$, showing that, on an individual basis, the size of onsets and resections were robust against the choice of parcellation schemes. When comparing the count of ALO regions, the sample sizes in each comparison varies between 52 and 56 as the number of subjects with regions appearing over the consensus threshold can vary. The parcellation scheme where the highest number of subjects have a consensus threshold is Lausanne-120 as the regions are larger in this scheme and therefore there is more chance of recurring regions, even if the onset channel itself varies across seizures.  

\end{document}